\documentstyle[12pt,fleqn]{article}
\def\beeq{\begin{equation}}
\def\eneq{\end{equation}}
\def\beeqa{\begin{eqnarray}}
\def\eneqa{\end{eqnarray}}

\addtocounter{section}{-1}
\begin{document}

\begin{center}

\vspace{2cm}

{\large {\bf {
Mechanism of magnetism in stacked nanographite:\\
Theoretical study
} } }

\vspace{1cm}

{\rm Kikuo Harigaya\footnote[1]{E-mail address: 
\verb+harigaya@etl.go.jp+; URL: 
\verb+http://www.etl.go.jp/+\~{}\verb+harigaya/+;
Address after April 2001: National Institute of
Advanced Industrial Science and Technology (AIST), 
Tsukuba 305, Japan}}

\vspace{1cm}

{\sl Physical Science Division,
Electrotechnical Laboratory,\\ 
Umezono 1-1-4, Tsukuba 305-8568, Japan}\footnote[2]{Corresponding address}\\
{\sl National Institute of Materials and Chemical Research,\\ 
Higashi 1-1, Tsukuba 305-8565, Japan}\\
{\sl Kanazawa Institute of Technology,\\
Ohgigaoka 7-1, Nonoichi 921-8501, Japan}

\end{center}

\vspace{1cm}

\noindent
{\bf Abstract}\\
Nanographite systems, where graphene sheets of the orders of 
the nanometer size are stacked, show novel magnetic properties, 
such as, spin-glass like behaviors and the change of ESR line 
widths in the course of gas adsorptions.  We theoretically 
investigate stacking effects in the zigzag nanographite sheets 
by using a tight binding model with the Hubbard-like onsite 
interactions.  We find a remarkable difference in the magnetic 
properties between the simple A-A and A-B type stackings. 
For the simple stacking, there are not magnetic solutions.
For the A-B stacking, we find antiferromagnetic solutions 
for strong onsite repulsions.  The local magnetic moments 
tend to exist at the edge sites in each layer due to the 
large amplitude of wavefunctions at these sites. 
Relations with experiments are discussed.


\pagebreak

\section{Introduction}

Nanographite systems, where graphene sheets of the orders of 
the nanometer size are stacked, show novel magnetic properties, 
such as, spin-glass like behaviors [1], and the change of 
ESR line widths while gas adsorptions [2]. Recently, it 
has been found [3] that magnetic interactions increase with the 
decrease of the interlayer distance while water molecules 
are attached physically.  Here, the change of the inter-layer 
interactions has been anticipated experimentally, but 
theoretical studies have not been reported yet.

In this paper, we theoretically consider the stacking 
effects in the zigzag nanographite sheets [4-6] by using a tight 
binding model with the Hubbard-like onsite interactions $U$. 
In the papers [4-6], the one dimensional graphite ribbons
have been investigated.  In this paper, we assume that
each graphite sheet has a hexagonal shape with zigzag edges.
Such the shape geometry has been used in the semi-empirical 
study of fluorine doped graphite nanoclusters [7], too.
The two stacking types, namely the A-A and A-B types, shown 
in Fig. 1 are considered in the model. We assume small interlayer 
interactions $t_1$ where two carbon atoms are adjacent between 
neighboring layers. The circles in Fig. 1 (a) (namely, 
nanographite {\bf a}) and Fig. 1 (c) (nanographite {\bf c})
show the sites with the interaction $t_1$, and the 
interaction $t_1$ is considered at all the sites in 
Fig. 1 (b) (nanographite {\bf b}).  Such the interactions
preserve the beautiful bipartite property seen in the single 
hexagonal layer.

The main finding of this paper is a remarkable difference 
in the magnetic properties between the simple A-A and A-B 
stackings.  For the simple stacking, we have not found
magnetic solutions, because the presence of local magnetic 
moments is suppressed at carbons.  For the A-B stacking, 
we have found antiferromagnetic solutions for $U>2t$, $t$ 
being the hopping integral in a layer. The local magnetic 
moments tend to exist at the edge sites in each layer due 
to the large amplitude of wavefunctions at these sites. 
Relations with experiments are discussed extensively.

In Sec II, we explain our model.  Sections III and IV are 
devoted to the total magnetic moment per layer, and
the local magnetic polarization per site, respectively.
In Sec. V, we discuss the local density of states
at the edge carbon atoms.  This paper is closed
with summary in Sec. VI.

\section{Model}

We study the following model with hopping integrals between
orbitals of carbon atoms and onsite strong repulsions of
the Hubbard type:
\beeqa
H &=& -t \sum_{\langle i,j \rangle: {\rm intralayer}} \sum_\sigma
(c_{i,\sigma}^\dagger c_{j,\sigma} + {\rm h.c.}) \nonumber \\
&-& t_1 \sum_{\langle i,j \rangle: {\rm interlayer}} \sum_\sigma
(c_{i,\sigma}^\dagger c_{j,\sigma} + {\rm h.c.}) \nonumber \\
&+& U \sum_i n_{i,\uparrow} n_{i,\downarrow},
\eneqa
where $n_i = c_{i,\sigma}^\dagger c_{i,\sigma}$ for 
$\sigma = \uparrow$ and $\downarrow$; $c_{i,\sigma}$ is
an annihilation operator of an electron at the $i$th site
with spin $\sigma$; the sum of the first line is taken
over the nearest neighbor pairs $\langle i,j \rangle$
in a single layer of the nanographite; the sum of the
second line is taken over sites where the distance 
between two positions of the neighboring layers is
shortest; $t_1$ is the strength of the weak hopping 
interaction between neighboring layers; the positions 
of $t_1$ are shown by the filled circles in Fig. 1 (a) 
(nanographite {\bf a}) and Fig. 1 (c) (nanographite {\bf c}); 
the interaction $t_1$ is considered at all the
sites in Fig. 1 (b) (nanographite {\bf b}); and the last 
term of the hamiltonian is the strong onsite repulsion 
with the strength $U$.

The finite size system is diagonalized numerically,
and we obtain two kinds of solutions.  One of them
is a nonmagnetic solution where up and down spin 
electrons are not polarized in each layer.  This 
kind of solutions can be found in weak $U$ cases.  
The other kind of solutions is an antiferromagnetic 
solution, where the number of up spin electrons is 
larger than that of down spin electrons in the first 
layer, the number of down spin electrons is larger 
than that of the up spin electrons in the second 
layer, and so on.  This kind of solution is realized 
in strong $U$ regions.  There will be cases of 
incommensurate spin density waves, but we have not 
obtained such kinds of solutions by choosing initial 
magnetic ordered states, which are commensurate 
with the one dimensional lattice in the stacking 
direction, at the first stage of the numerical 
iteration process.  The present author has discussed
the antiferromagnetism in C$_{60}$ polymers [8].  
The same technique used in Ref. [8] is effective 
in this paper, too.

The parameters are changed within $0 \leq t_1 \leq 0.5t$
and $0 \leq U \leq 10t$.  The realistic value of $t_1$
is estimated to be about $0.1t$ at most, but we change
this parameter for more extended regions in order to 
look at the behaviors of solutions in detail.  All of the
quantities of the energy dimension are reported using
the unit $t$ ($\sim 2.0 - 3.0$ eV).

\section{Magnetic moment per layer}

First, we consider the total magnetic moment per layer 
for the nanographites {\bf a} and {\bf b}.
Figure 2 shows the absolute value of the total magnetic 
moment per layer as functions of $t_1$ and $U$.  Figures 2
(a) and (b) are for the nanographite {\bf a},
and Fig. 2 (c) is for the nanographite {\bf b}.
See the figure caption for the value of $U$ of each plot.
Figure 2 (a) shows the overall variations of the magnetic 
moment.  When $U$ is small, there appears a finite
magnetic moment for the values of $t_1$ larger than the
threshold of the phase transition.  At $U=2.5t$,
the magnetic moment changes like a function of a parabola
with respect to $t_1$.  The magnetic moment decreases for 
larger $U$:  $U = 3.0t$, $5.0t$, and $10.0t$.  This is
due to the strong singlet correlation at the bonds
$t_1$ with respect to the change of the Heisenberg coupling 
between the neighboring layers as $t_1^2/U$.  Figure 2 (b) 
shows the details around the phase transition for 
$1.8t \leq U \leq 2.3t$.  As increasing $U$, the magnitude 
of the magnetization increases.  The magnetic moment is 
zero at the smaller $t_1$ region for $U = 1.8t$, $1.9t$, 
$2.0t$, and $2.1t$.  The magnetic moment is zero only at 
$t_1 = 0$ for $U=2.2t$ and $2.3t$.  We can understand the 
parabolic curves as a change due to the Heisenberg coupling 
proportional to $t_1^2/U$.  The antiferromagnetic solutions 
really exist for larger $U$ regions in the A-B stacking case.
On the other hand, Fig. 2 (c) shows the magnetic moment
for the simple stacking.  There is not magnetization
for $0 \leq t_1 \leq 0.5t$ and $0 \leq U \leq 10t$.
This is a remarkable difference between the simple
A-A and A-B stackings, and is a new finding of this
paper.  There is not any clear evidence which stacking
is realized experimentally [3].  However, we believe that
the A-B stacking should exist in nanographite systems
because the exotic magnetisms have been observed
in recent experiments [1-3].  The increase of the 
magnetic interaction while attachment of water molecules [3] 
is induced by the decrease of the interlayer distance,
which enhances the interaction strength $t_1$.

Next, Fig. 3 shows the absolute value of the magnetic
moment per layer for the nanographite {\bf c} with
A-B stackings.  The carbon number in a layer is 54,
and is more than twice as large as that of the 
nanographite {\bf a}.  Figure 3 (a) shows the change
of the magnetic moment for wide parameter regions,
and Fig. 3 (b) displays the numerical data around the
phase transition.  The overall behaviors in the smaller
and larger $U$ seem similar to those of Fig. 2.
The characteristic value of $U$ decreases from that
of Fig. 2.  For example, the curve becomes parabolic
for $U \geq 2.2t$ in Fig. 2 (b), and it becomes
parabolic for $U \geq 2.0t$ in Fig. 3 (b).  Such the 
quantitative difference is due to the effects of the 
larger system size.

\section{Local magnetic polarization in a layer}

In order to give further insights into mechanism
of the magnetism, we will look at the local magnetic
moments which depend on the carbon sites in each layer.
We particularly pay attention to the local magnetism
near the edge sites of the nanographite [4-6].

Figure 4 shows the local magnetic moment at the edge 
sites of the nanographite {\bf a}.  The values of $U$ 
are $U=2.1t$ and $2.5t$ in Figs. 4 (a) and (b), 
respectively.  In the former case, there is a point
of the phase transition near $t_1 = 0.25t$, and there 
appears a finite magnetization for nonzero $t_1$ 
in the latter case.  The filled squares, open squares, 
and filled circles show the results at sites A, B, and C, 
respectively.  Due to the symmetry, the magnetic moments 
at the sites A', B', and C' of Fig. 1 (a) are equal 
to those of the sites A, B, and C.  Figure 5 displays the
similar plots for the nanographite {\bf c}.  The parameters
are $U = 1.8t$ and $2.5t$ in Figs. 5 (a) and (b),
respectively.  We see that the local magnetic
moment is negative along the edge A-A' in Fig. 1 (a),
and also along the edge A-B-A' in Fig. 1 (c).
The local magnetic moment is positive along the
neighboring edges: namely, the edges B-C and B'-C'
in Fig. 1 (a), and the edges C-D-E and C'-D'-E' in Fig. 1 (c).
Such the positive and negative alternations of magnetic
moments are seen in both calculations of the nanographites
{\bf a} and {\bf c}.  Because there are strong amplitudes 
of wavefunctions at the zigzag edge sites [4-6], the 
local moments near these edge carbon atoms tend to 
become larger.  In the A-B stackings, there is not
an interlayer interaction $t_1$ at the edge sites,
and this gives rise to the finite magnetic moment
per layer which has been discussed in the previous 
section.  On the other hand, the local magnetic moment
and also the magnetic moment per layer do not appear
in the simple A-A stacking case namely the nanographite
{\bf b} of Fig. 1 (b), owing to the interlayer 
interactions $t_1$ which are present between all the 
nearest carbon atoms of neighboring layers.  This 
difference is the origin of the fact that there is
not magnetization in the simple stacking case
reported in Fig. 2 (c).

In the band calculations of the stacked nanographite
ribbons [9], the strong hybridization between edge 
states occurs in the A-A stacking case.  Such the 
hybridization is weak in the A-B stacking case.  The 
strong localization of wavefunctions at the edge 
carbon sites persists in the band calculations for 
systems with the A-B stacking [9], and this property 
agrees with the present result.

\section{Density of states}

In this section, we discuss the local density of states
at the edge sites.  The wavefunctions of electrons with
up and down spins are projected on the edge sites of the
nanographite {\bf a} and {\bf c}.  The local density of
states is reported together with the total density of states.

Figure 6 shows the density of states of the nanographite 
{\bf a}, and Fig. 7 displays the result of the nanographite
{\bf c}.  The total density of states per layer and per spin
is shown by the bold line.  The local density of states
at the edge sites is shown by the thin and dashed lines
for the up and down spins, respectively.  The up and down
splitting typical to the antiferromagnetism is seen
in both figures.  Because the number of edge sites is 
one third of that of the total carbon atoms in the 
nanographite {\bf a}, the areas between the lines and
the horizontal axis have such the relative ratios.
In the nanographite {\bf c}, the portion of the edge
sites with respect to the total carbon number becomes 
smaller.  Therefore, the relative area below the thin 
and dashed lines becomes smaller in Fig. 7.  In one 
dimensional graphite ribbons [4-6], there appears
a strong peak due to the localized edge states at the
Fermi energy.  This is seen in the non-interacting case.
With interactions taken into account, such the edge
states split into bonding (occupied) and antibonding
(unoccupied) states.  This fact will be one of the reasons
why such the strong peak is not observed in Figs. 6 and 7.
Also, in the present case, the edge sites do not make 
a one dimensional lattice and each layer has a finite
spatial dimension.  Such the difference will be the
second reason of the absence of the strong peak.

\section{Summary}

In summary, we have theoretically investigated the stacking 
effects in the zigzag nanographite sheets.  We have found a 
remarkable difference in the magnetic properties between the 
simple A-A and A-B type stackings.  For the simple stacking, 
there are not magnetic solutions.  For the A-B stacking, we 
find antiferromagnetic solutions for strong onsite repulsions.  
The local magnetic moments exist at the edge sites due to the 
large amplitude of wavefunctions at the zigzag edge sites.
The A-B type stacking is favorable in order that the
exotic magnetism is observed in nanographite systems.

\begin{flushleft}
{\bf Acknowledgements}
\end{flushleft}

\noindent
The author is grateful for interesting discussion with
T. Enoki, T. Kawatsu, T. Ohshima, Y. Miyamoto, K. Kusakabe, 
K. Nakada, K. Wakabayashi, and M. Igami.  Useful discussion 
with the members of Condensed Matter Theory Group
(\verb+http://www.etl.go.jp/+\~{}\verb+theory/+),
Electrotechnical Laboratory is acknowledged, too.

\pagebreak
\begin{flushleft}
{\bf References}
\end{flushleft}

\noindent
$[1]$ Y. Shibayama, H. Sato, T. Enoki, and M. Endo, 
Phys. Rev. Lett. {\bf 84}, 1744 (2000).\\
$[2]$ N. Kobayashi, T. Enoki, C. Ishii, K. Kaneko, and M. Endo,
J. Chem. Phys. {\bf 109}, 1983 (1998).\\
$[3]$ N. Kawatsu, H. Sato, T. Enoki, M. Endo, 
R. Kobori, S. Maruyama, and K. Kaneko,
Meeting Abstracts of the Physical Society of Japan
{\bf 55} Issue 1, 717 (2000).\\
$[4]$ M. Fujita, K. Wakabayashi, K. Nakada, and K. Kusakabe,
J. Phys. Soc. Jpn. {\bf 65}, 1920 (1996).\\
$[5]$ M. Fujita, M. Igami, and K. Nakada,
J. Phys. Soc. Jpn. {\bf 66}, 1864 (1997).\\
$[6]$ K. Nakada, M. Fujita, G. Dresselhaus, and M. S. Dresselhaus,
Phys. Rev. B {\bf 54}, 17954 (1996).\\
$[7]$ R. Saito, M. Yagi, T. Kimura, G. Dresselhaus, and
M. S. Dresselhaus, J. Phys. Chem. Solids {\bf 60},
715 (1999).\\
$[8]$ K. Harigaya, Phys. Rev. B {\bf 53}, R4197 (1996).\\
$[9]$ Y. Miyamoto, K. Nakada, and M. Fujita,
Phys. Rev. B {\bf 59}, 9858 (1999).\\

\pagebreak
\begin{flushleft}
{\bf Figure Captions}
\end{flushleft}

\mbox{}

\noindent
Fig. 1. Stacked nanographite with zigzag edges.
The bold and thin lines show the first and second
layers, respectively.  The stacking is the A-B type 
in (a) (nanographite {\bf a}) and (c) (nanographite
{\bf c}), and it is the simple A-A type in (b)
(nanographite {\bf b}).  There are 24 carbon atoms 
in one layer in (a) and (b), and there are 54 atoms 
in one layer in (c).  The filled circles in (a) and (c) 
show lattice positions with small interlayer 
interactions $t_1$, and the bold symbols indicate 
some of edge sites in the first layer.  The sites, 
A (B, C ...) and A' (B', C' ...), are symmetrically 
equivalent, respectively.

\mbox{}

\noindent
Fig. 2.  The magnitude of the total magnetic moment
per layer as functions of $t_1$ and $U$.  Figures 
(a) and (b) are for the nanographite {\bf a},
and Fig. (c) is for the nanographite {\bf b}.
In (a), the values of $U$ are $U=1.5t$ (filled
squares), $2.0t$ (open squares), $2.5t$ (filled circles),
$3.0t$ (open circles), $5.0t$ (filled triangles),
and $10.0t$ (open triangles), respectively. 
Figure (b) shows the details around the phase 
transition: the values of $U$ are $U=1.8t$ (filled
squares), $1.9t$ (open squares), $2.0t$ (filled circles),
$2.1t$ (open circles), $2.2t$ (filled triangles),
and $2.3t$ (open triangles), respectively. 
In (c), the magnetic moment is zero for 
$0 \leq U \leq 10t$.

\mbox{}

\noindent
Fig. 3.  The magnitude of the total magnetic moment
per layer as functions of $t_1$ and $U$ for the 
nanographite {\bf c}.
In (a), the values of $U$ are $U=1.0t$ (filled
squares), $1.5t$ (open squares), $2.0t$ (filled circles),
$2.5t$ (open circles), $5.0t$ (filled triangles),
and $7.0t$ (open triangles), respectively. 
Figure (b) shows the details around the phase 
transition: the values of $U$ are $U=1.2t$ (filled
squares), $1.4t$ (open squares), $1.6t$ (filled circles),
$1.8t$ (open circles), $2.0t$ (filled triangles),
and $2.2t$ (open triangles), respectively.

\mbox{}

\noindent
Fig. 4.  Local magnetic moment at the edge sites
of the nanographite {\bf a}.  The values of $U$ 
are $U=2.1t$ in (a), and $2.5t$ in (b).
The filled squares, open squares, and filled
circles show the results at sites A, B, and C,
respectively.

\mbox{}

\noindent
Fig. 5.  Local magnetic moment at the edge sites
of the nanographite {\bf c}.  The values of $U$ 
are $U=1.8t$ in (a), and $2.5t$ in (b).
The filled squares, open squares, filled circles,
open circles, and filled triangles show the results 
at sites A, B, C, D, and E, respectively.

\mbox{}

\noindent
Fig. 6.  Density of states per layer of the 
nanographite {\bf a}.  The parameters are
$t_1 = 0.4t$ and $U=2.1t$.  The bold line shows
the density of states over 24 carbon atoms per layer
and per spin.  The thin and dashed lines indicate
the density of states over the eight edge sites
in a layer for the up and down spins, respectively.

\mbox{}

\noindent
Fig. 7.  Density of states per layer of the 
nanographite {\bf c}.  The parameters are
$t_1 = 0.4t$ and $U=1.8t$.  The bold line shows
the density of states over 54 carbon atoms per layer
and per spin.  The thin and dashed lines indicate
the density of states over the twelve edge sites
in a layer for the up and down spins, respectively.

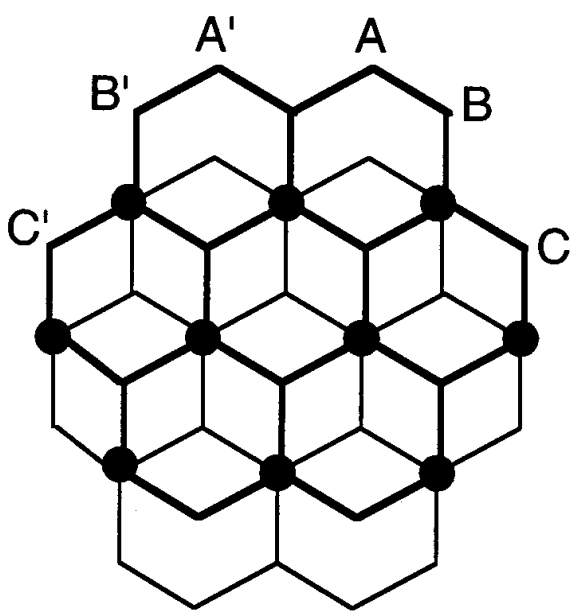
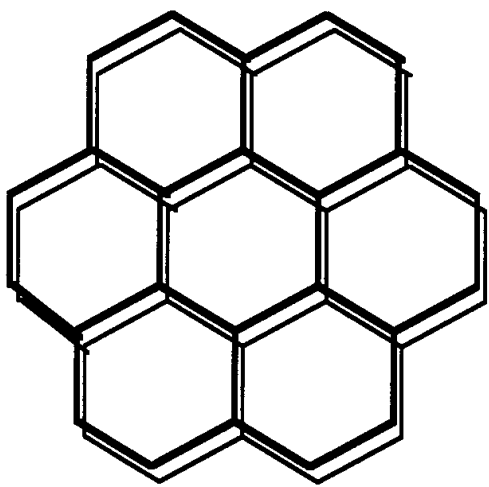
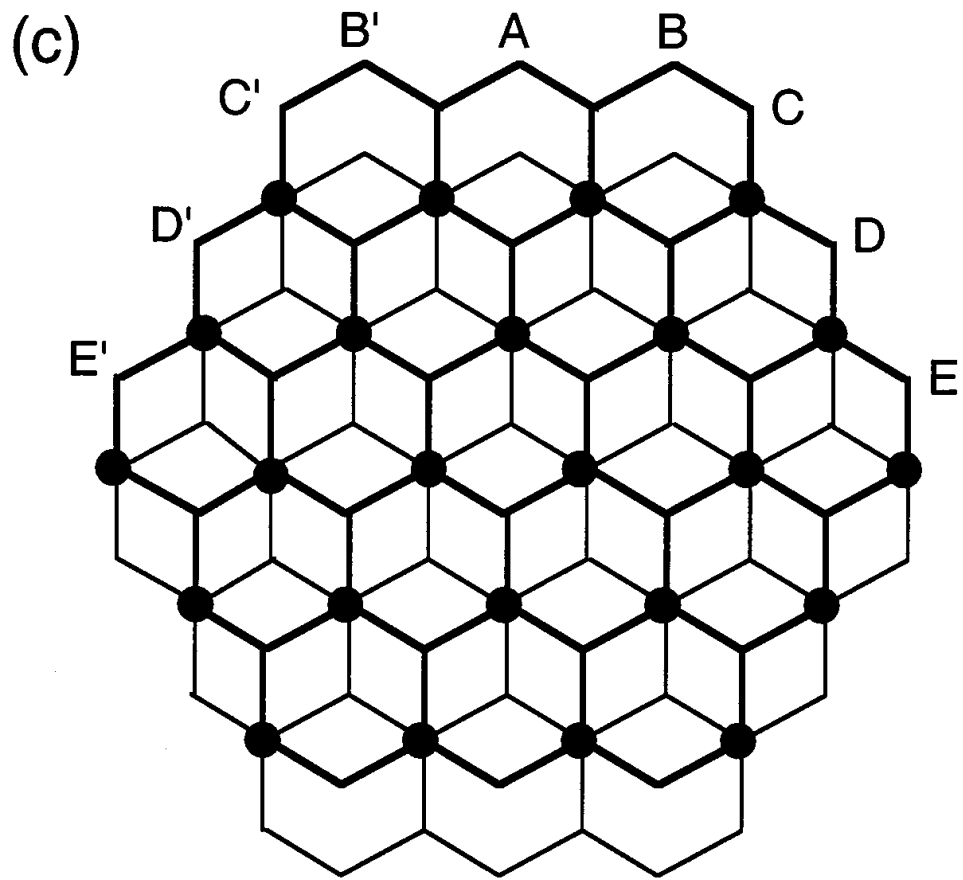
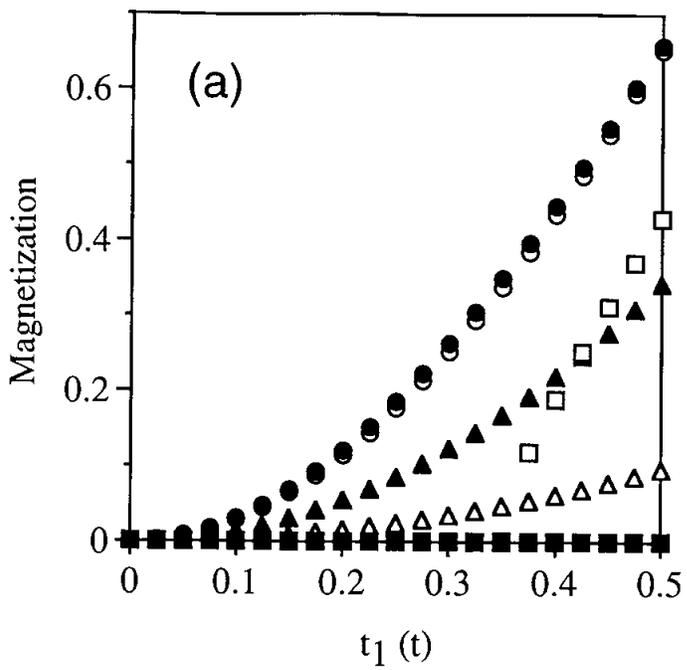
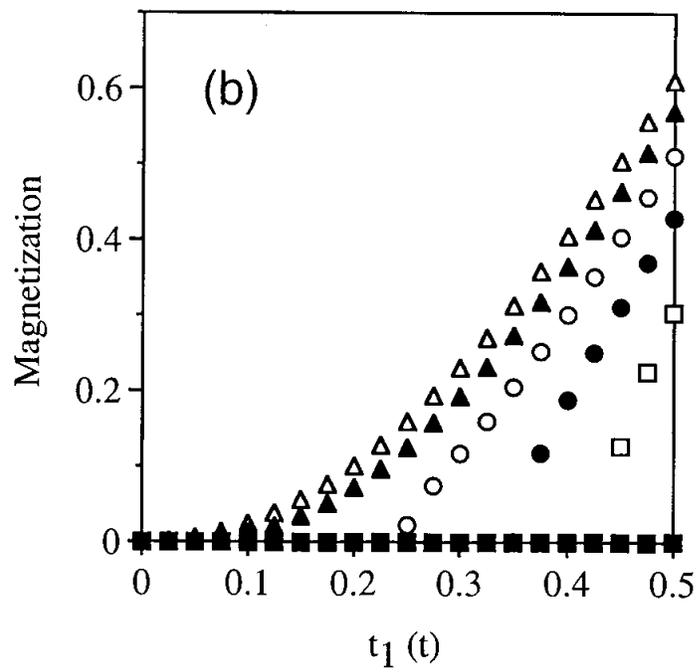
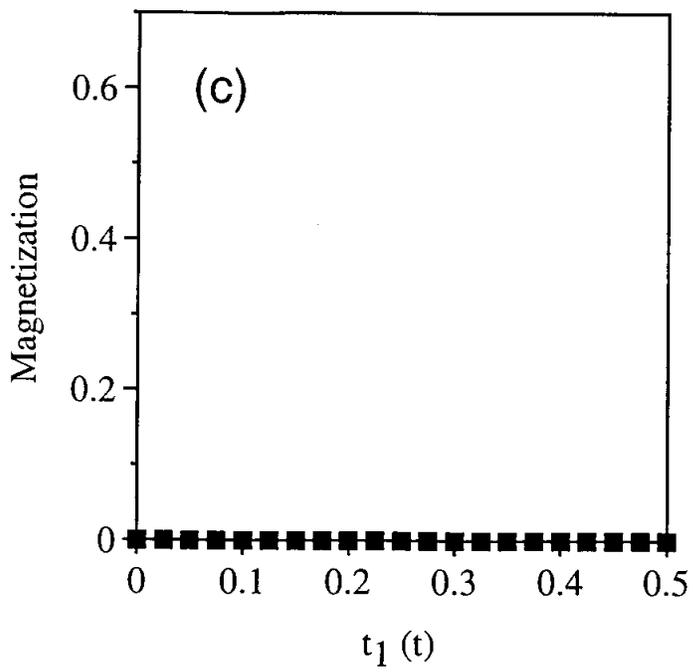
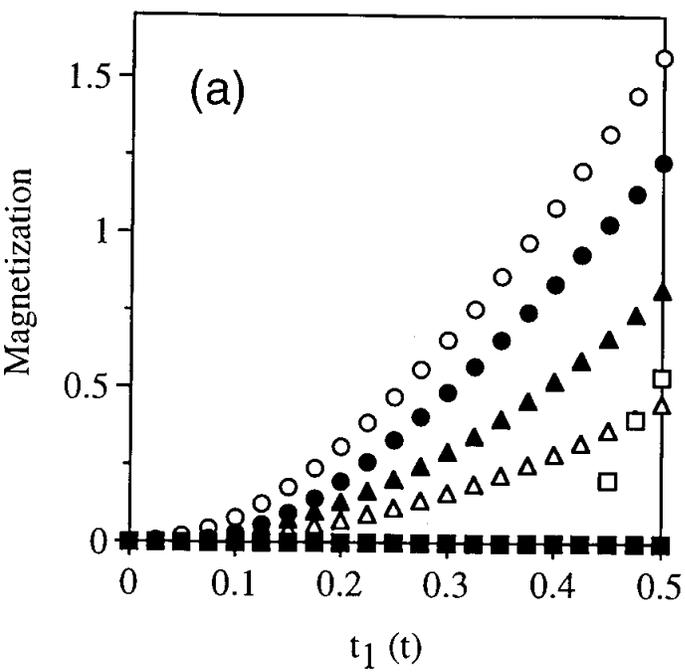
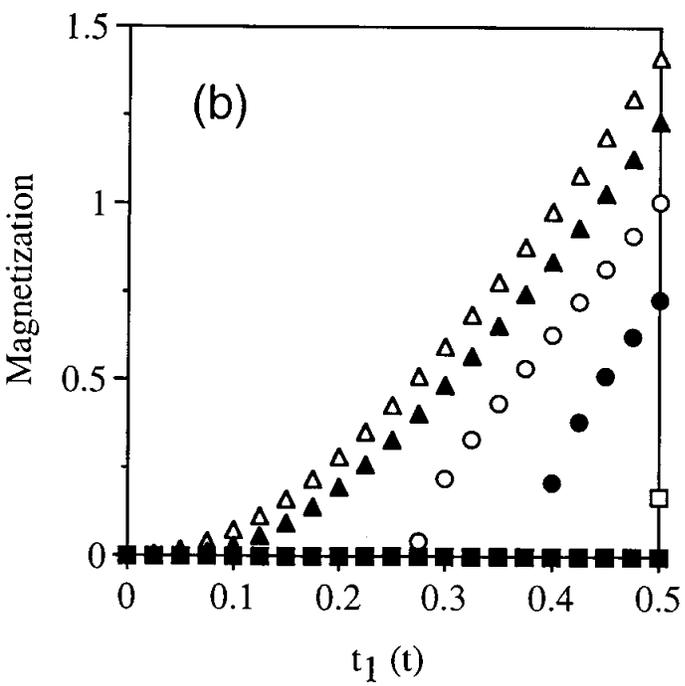
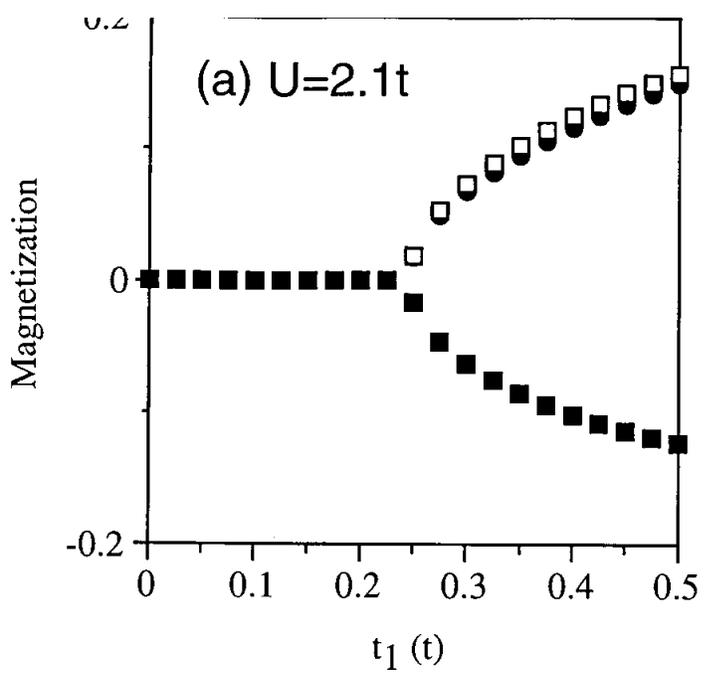
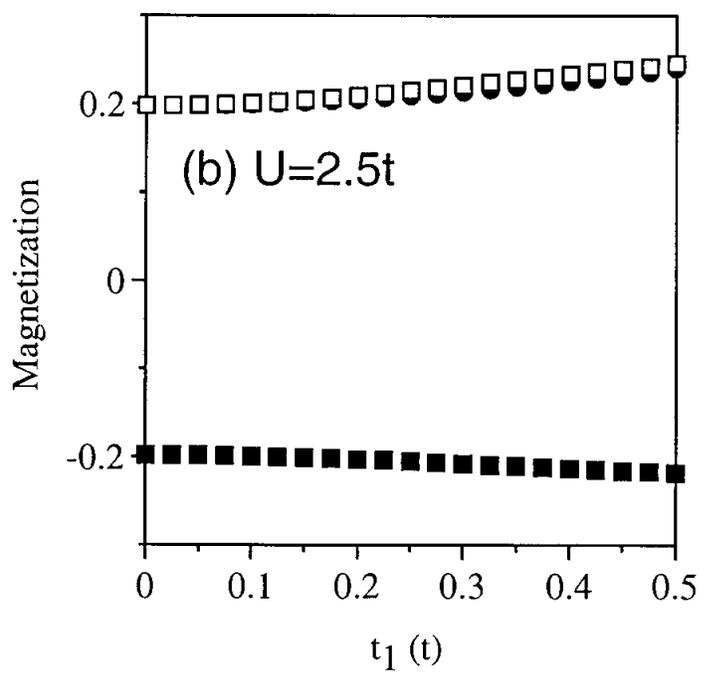
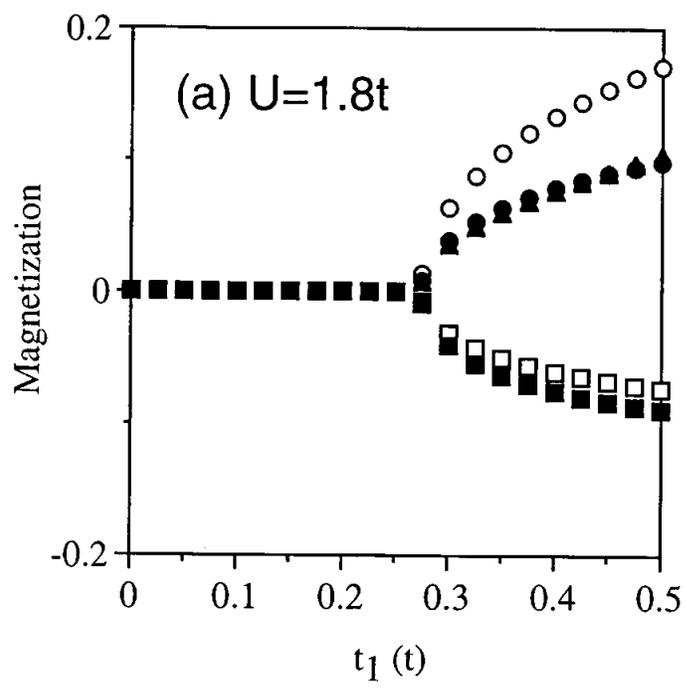
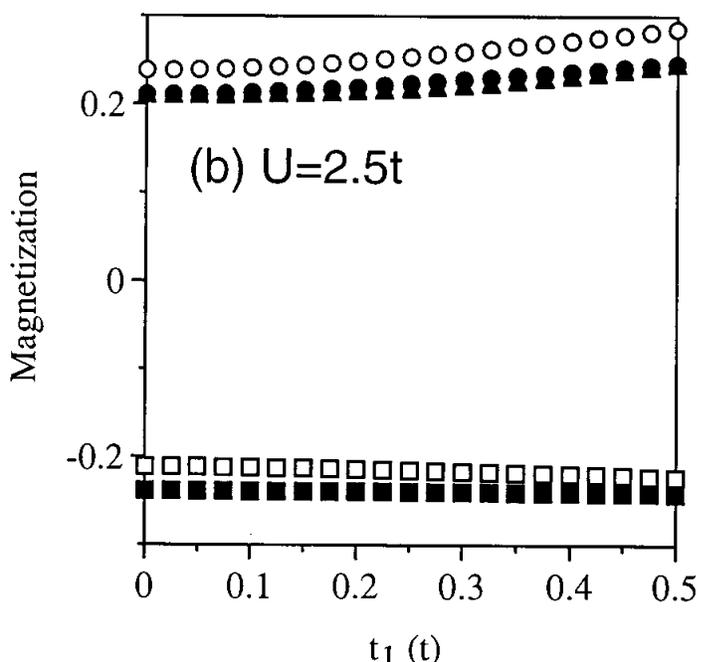
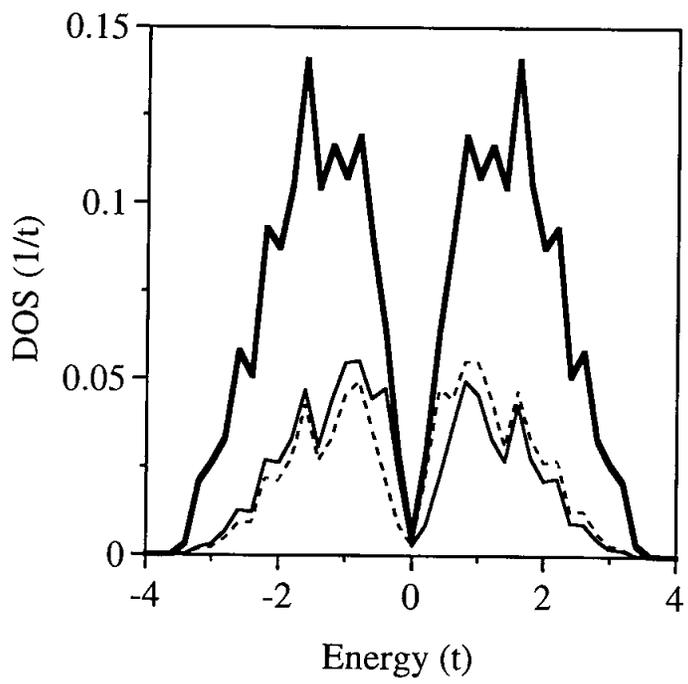
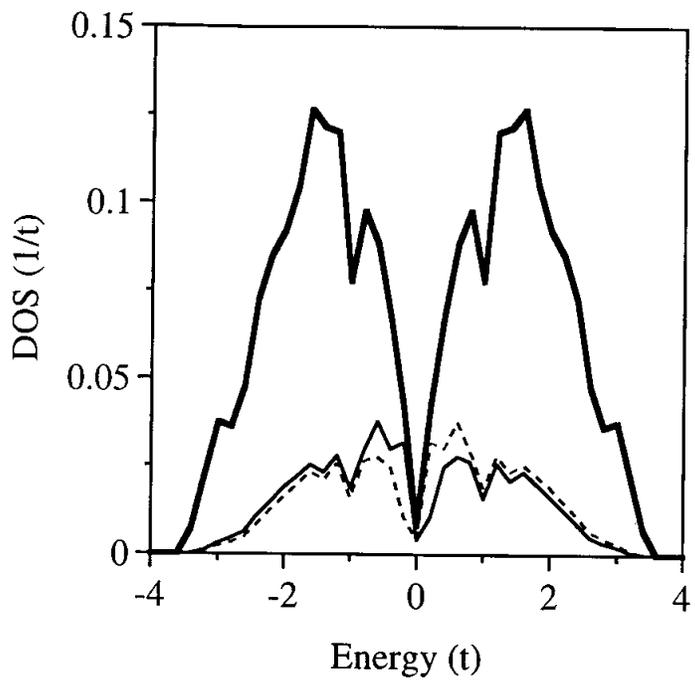
\end{document}